\documentclass[reprint, aps, prb, showpacs, superscriptaddress]{revtex4-1}
\usepackage{amsmath}   
\usepackage{graphicx}
\usepackage[utf8]{inputenc}
\usepackage[english]{babel}

\usepackage{hyperref}

\begin{document}
\title{Cu nuclear magnetic resonance study of charge and spin stripe order in La$_{1.875}$Ba$_{0.125}$CuO$_4$}

\author{D. Pelc}
\affiliation{Department of Physics, Faculty of Science, University of Zagreb, Bijeni\v{c}ka 32, HR-10000, Zagreb, Croatia}

\author{H.-J. Grafe}
\affiliation{IFW Dresden, Institute for Solid State Research, P.O. Box 270116, D-01171 Dresden, Germany}

\author{G. D. Gu}
\affiliation{Condensed Matter Physics and Materials Science Department, Brookhaven National Laboratory, Upton, NY 11973-5000, USA}

\author{M. Požek}
\email{Correspondence to: mpozek@phy.hr}
\affiliation{Department of Physics, Faculty of Science, University of Zagreb, Bijeni\v{c}ka 32, HR-10000, Zagreb, Croatia} 

\begin{abstract}

We present a Cu nuclear magnetic/quadrupole resonance study of the charge stripe ordered phase of LBCO, with detection of previously unobserved ('wiped-out') signal. We show that spin-spin and spin-lattice relaxation rates are strongly enhanced in the charge ordered phase, explaining the apparent signal decrease in earlier investigations. The enhancement is caused by magnetic, rather than charge fluctuations, conclusively confirming the long-suspected assumption that spin fluctuations are responsible for the wipeout effect. Observation of the full Cu signal enables insight into the spin and charge dynamics of the stripe-ordered phase, and measurements in external magnetic fields provide information on the nature and suppression of spin fluctuations associated with charge order. We find glassy spin dynamics, in agreement with previous work, and incommensurate static charge order with charge modulation amplitude similar to other cuprate compounds, suggesting that the amplitude of charge stripes is universal in the cuprates. 

\end{abstract}


\maketitle

\section{Introduction}

Since their discovery, cuprate superconductors have been extensively studied by nuclear magnetic resonance (NMR), with many important insights. Yet some issues remain unresolved, and the technique is still capable of providing fresh information on the local electronic physics in the cuprates. Recently, charge and spin stripe order has emerged as one of the crucial ingredients of the cuprate phenomenology\cite{LBCOneutron,LBCO2D,COxray1,COxray2,COxray3,COxray4,COxray5}, and NMR in these ordered phases has once more come into focus. In the yttrium-123 (YBCO) family, copper NMR has provided direct evidence of a commensurate charge stripe ordering at high magnetic fields\cite{YBCONMR1,YBCONMR2,YBCONMR3}, while lanthanum NMR in the 214 family (La$_{2-x}$Sr$_x$CuO$_4$ - LSCO, La$_{2-x}$Ba$_x$CuO$_4$ - LBCO La$_{2-x-y}$Eu$_y$Sr$_x$CuO$_4$ - LESCO and similar) detects the influence of stripe-related fluctuations\cite{LSCOLaNMR1,LSCOLaNMR2,LBCOLaNMR,stripe_glass,LESCOLaNMR}. However, charge stripes in the 214 family have never been directly studied by in-plane copper NMR/NQR, due to an effect known as wipeout. It has been known for a long time that in stripe-ordered lanthanum cuprates such as LBCO and LESCO the copper NMR signal diminishes -- is wiped out -- in the charge-ordered phase\cite{wipeout1,wipeout2,wipeout3,stripe_glass,Owipeout,wipeout4,LESCOmi}. The fraction of wiped-out nuclei has been phenomenologically related to the charge stripe order parameter\cite{wipeout1,wipeout2}, but without access to the disappeared signal the true nature of wipeout is difficult to ascertain and alternative explanations have been put forward\cite{wipeout4,wipeoutspin}. Also, a marked difference between wipeout fractions in powder and single crystals is observed\cite{wipeout1,stripe_glass,wipeout_zn,wipeout3}, as well as differences between Cu and O NMR and NQR\cite{YBCONMR1,YBCONMR3,Owipeout}. Copper wipeout is also seen in Ni-codoped cuprates\cite{wipeout_zn} and La-based cuprates at doping close to the insulator-superconductor boundary\cite{wipeout_UD}, indicating a more general relationship with spin, rather than charge fluctuations\cite{wipeoutspin}. Yet important insight into the stripes would be gained from Cu NMR/NQR if the signal could be detected, since copper nuclei have a quadrupolar moment and the signals are thus sensitive to charge effects. A direct comparison to e.g. Cu NMR in YBCO could be made, highlighting the differences and similarities of charge stripes in these representative cuprates. 

Our purpose in this work is twofold: to present an experimental methodology which enables the detection of the previously unseen Cu nuclei -- establishing beyond doubt that wipeout is caused by spin fluctuations -- and to apply the methodology to provide insight into the ordered phase of the archetypal striped 214-family cuprate LBCO. We show that the NQR spectra and relaxation times of the previously unobserved Cu nuclei in LBCO are indeed highly sensitive to the charge order, enabling us to draw conclusions about the structure and dynamics of the stripes and compare them to other cuprates. Furthermore, the experimental modifications we employed for studying the quickly relaxing Cu NQR signals here can be useful in a wide range of systems previously unavailable for NMR due to strong wipeout. 

The paper is organized as follows: in Sec. II we expound the experimental methodology used in detecting the wiped-out Cu signal, in Sec. III we present the results of Cu NQR on LBCO, including spectra and relaxation times in dependence on temperature; in Sec. IV we investigate the effect of external magnetic fields on the spin fluctuations causing wipeout; in Sec. V we discuss the results and we summarize in Sec. VI.

\section{Experimental}

In order to detect the quickly-relaxing 'wiped-out' nuclei in an NMR/NQR experiment, it is crucial to employ analogue signal processing electronics with recovery times as short as possible. A typical NMR/NQR experiment uses a sample coil which is part of a resonant LC circuit, a high-power radio-frequency (RF) amplifier to supply the pulses which move the spin system out of equilibrium, and a low-noise signal preamplifier as the first active signal processing stage. Pulse powers are usually tens of Watts, and the signal to be detected is many orders of magnitude smaller. Hence the primary concern is isolating the sensitive preamplifier from the pulses, and if the signal decays quickly, the preamplifier must be able to operate a very short time after being overloaded by the pulse itself. In this work we employed Miteq RF preamplifiers with overload recovery times as short as 1~$\mu$s, setting the lower limit for signal relaxation times. The second important issue is ringing of the LC resonant circuit after the pulses, which often completely obscures the signal for tens of $\mu$s. Ringing can be suppressed in a number of ways, with several active damping schemes envisaged previously\cite{damp1,damp2}. However, most of them do not operate quickly enough for our purposes, the only possibility thus being passive damping i.e. spoiling the Q-factor of the resonant circuit. Yet as the Q-factor decreases, so does the sensitivity of the circuit, implying that a price in signal intensity must be paid for measurements at short times. In our work this is not very problematic, since $^{63}$Cu is a sensitive NMR nucleus with high natural abundance and plenty of signal intensity. Also, the employed pulse sequence and phasing of the pulses can play an important role in suppressing the ringing. We employ a conventional Hahn spin echo sequence with double anti-ringing phase cycling, further decreasing the ringing problem. Finally, the spectrometer must be capable of fast measurements after pulses -- we use a commercial Tecmag Apollo NMR/NQR spectrometer with dead time below 1~$\mu$s. All the described experimental considerations enable us to perform a reliable NMR/NQR spin echo measurement with echo times as short as 2~$\mu$s, almost an order of magnitude faster than typical experiments. With state-of-the-art technology it would be difficult to improve on these times, and the only further possibility for detection of nuclei with spin-spin relaxation faster than 1~$\mu$s we are aware of is high-power continuous wave NMR\cite{fukushima}. Fortunately, it appears that in studying cuprates this is not necessary.  

Apart from the above-described modifications, we performed a rather standard NQR experiment with a single crystal sample of LBCO-1/8 extensively used and characterized in previous work\cite{LBCOneutron,LBCO2D,LBCOtransport}. The measurements were made in a variable-temperature inset of a 12~T Oxford superconducting magnet, with the sample mounted on a sapphire holder to minimize heating from currents induced in the (conductive) sample by the RF pulses used in NMR/NQR. Spin-lattice relaxation times were measured using a saturation recovery sequence with Hahn echo detection. 

\section{NQR results}

\begin{figure}
\centering
\includegraphics[width=84mm]{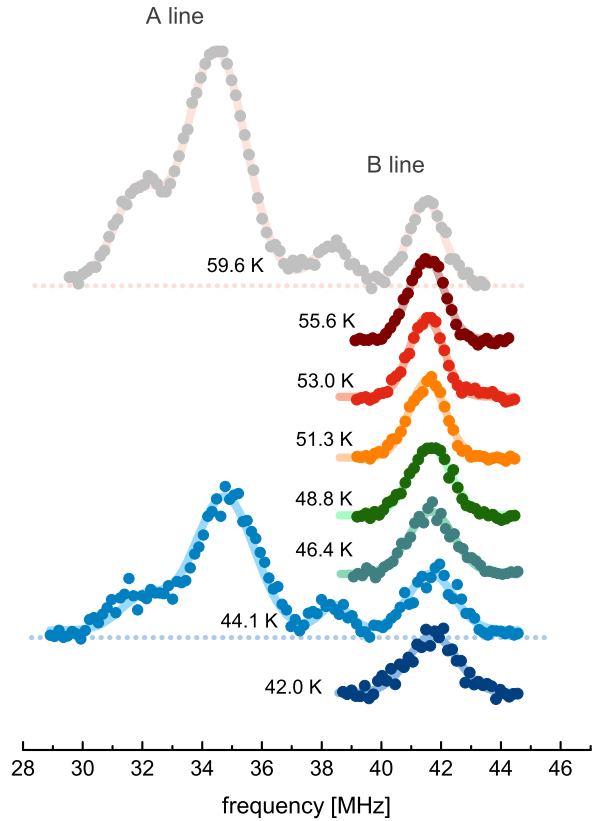}
\caption{\label{spektri} (color online) Cu NQR spectra of LBCO in dependence on temperature, measured with an echo time of 2.3~$\mu$s, normalized to their total areas and shifted vertically for clarity. Two sets of lines are observed, referred to as A and B lines. Both appear twice for the two isotopes $^{65}$Cu and $^{63}$Cu. Charge order sets in below $\sim 53$~K, where a progressive broadening of the lines is seen.  
}
\end{figure}

The basic information to be obtained from an NQR experiment is the shape of spectral lines. The NQR spectrum of LBCO above the charge-ordering temperature $T_{CO}$ is well-known\cite{stripe_glass}, and our results correspond exactly to previously published spectra (Fig. \ref{spektri} topmost spectrum). Since 214 cuprates are doped by ions close to their copper oxide planes, two distinct lines are always observed in copper NQR spectra. The high-frequency signal -- called the B line -- is from Cu nuclei close to the dopand ions (in our case Ba), while the low-frequency A line is from all other Cu nuclei. Two sets of A and B lines are observed for two Cu isotopes, $^{63}$Cu and $^{65}$Cu. A fortunate circumstance makes LBCO particularly suited for investigating the effects of charge order on the Cu NQR spectrum -- the B line is well separated from the A line (in contradistinction to other 214 cuprates such as LSCO and LESCO). This eliminates the need for complicated spectral deconvolution, simplifying the analysis considerably. Hence we concentrate on the evolution of the $^{63}$Cu B line with temperature in the charge-ordered phase (Fig. \ref{spektri}).

\begin{figure}
\centering
\includegraphics[width=87mm]{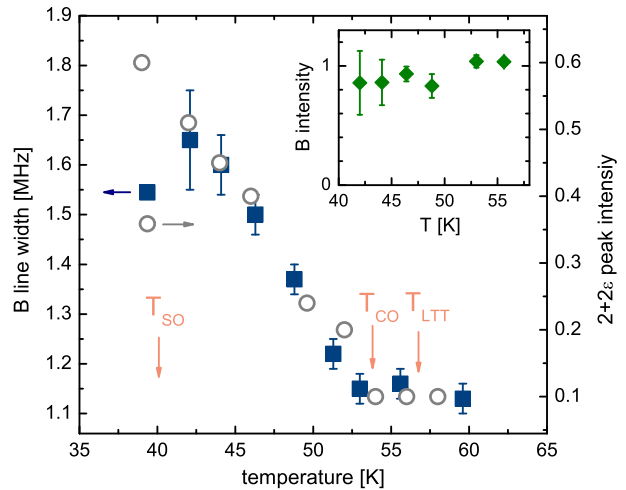}
\caption{\label{order} (color online) Width of the $^{63}$Cu NQR B line in dependence on temperature in the charge ordered phase, compared to a resonant X-ray measurement of the charge order peak intensity\cite{LBCOtransport}. The NQR linewidth is seen to correspond exactly to the charge order parameter, as one should expect for incommensurable stripes. The LTO-LTT structural transition at $T_{LTT} \approx 57$~K and the spin-ordering temperature $T_{SO} \approx 40$~K are from a neutron diffraction study\cite{LBCOneutron}. Inset shows the integrated B line signal intensity corrected for temperature and spin-spin relaxation, demonstrating that no significant decrease occurs below $T_{CO}$.
}
\end{figure}

Clearly the B line broadens significantly between 53~K and 42~K: a quantitative analysis using a simple Gaussian fit to the line is shown in Fig. \ref{order}. The linewidth can be compared to a hard X-ray measurement of the charge order peak\cite{LBCOtransport} which is proportional to the charge stripe order parameter (circles in Fig. \ref{order}). The B linewidth corresponds exactly to the X-ray data, and thus is a microscopic measure of the order parameter. This should be expected for an incommensurate charge ordering, and is also observed e.g. in YBCO at high temperatures\cite{YBCONMR3}. Within our experimental uncertainty, we see no commensurate order (which would result in line splitting, rather than broadening\cite{YBCONMR1}). The A line broadens as well, but its analysis is unreliable due to the overlap of $^{63}$Cu and $^{65}$Cu lines and their greater width. Also, in the charge-ordered phase the A line shape appears to deviate from simple Gaussian (appropriate at 60~K), indicating that a nontrivial spectral redistribution takes place. This is not surprising, considering the intrinsic charge disorder of 214 cuprates\cite{214disorder} and that the A line signal originates from Cu sites at different distances from the Ba impurities. We note that the structural transition at $\sim 56$~K and the associated phase separation cannot be the cause of the observed line broadening, since it is known from neutron\cite{LBCOneutron} and X-ray\cite{LBCOtransport} studies that the low-temperature tetragonal (LTT) phase completely replaces the low-temperature orthorhombic (LTO) phase within 2~K of the transition. 

\begin{figure}[b]
\centering
\includegraphics[width=79mm]{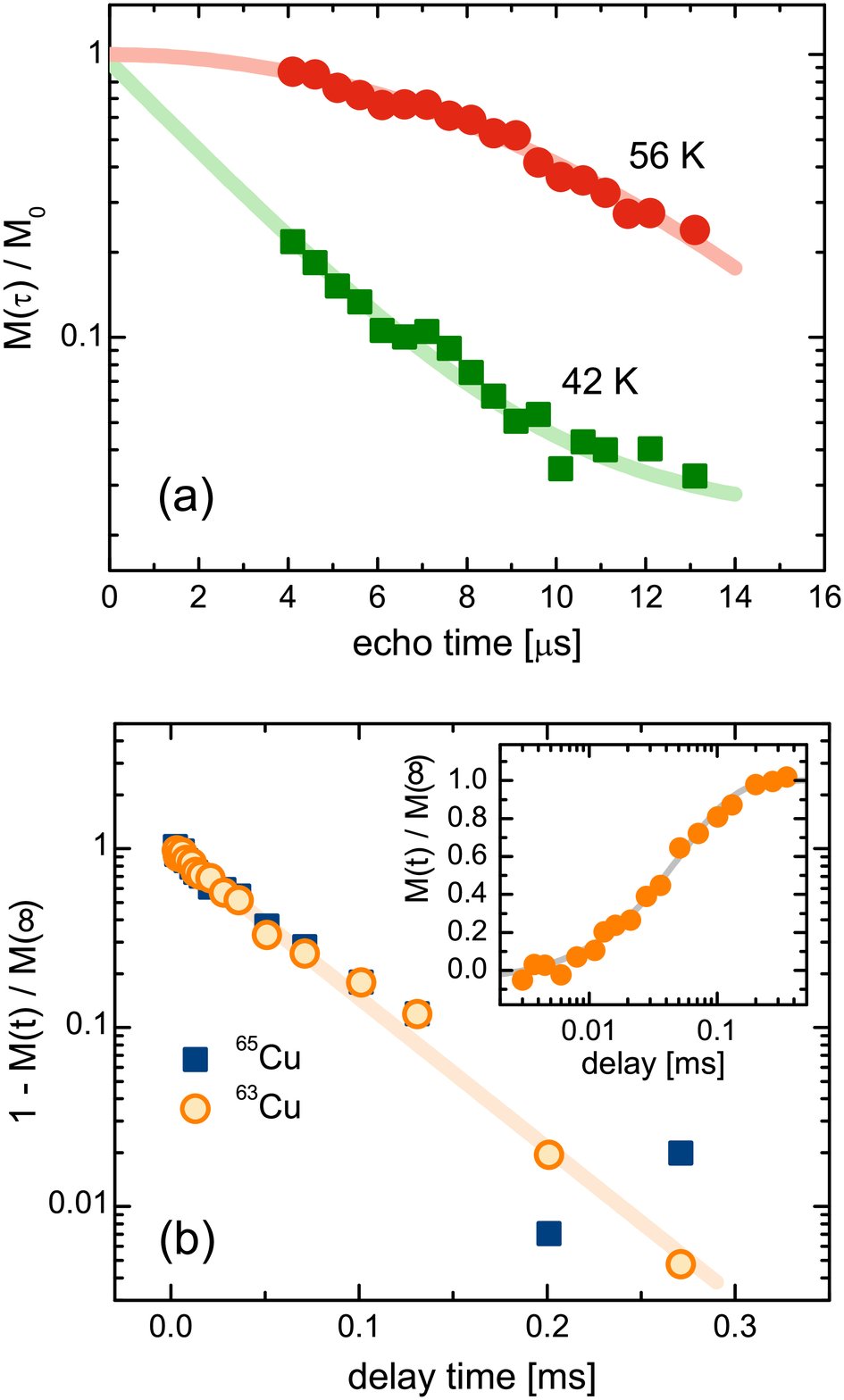}
\caption{\label{spins} (color online) (a) Spin-spin relaxation curves measured at the center of the B line at two representative temperatures. Both curves are normalized to $M_0 (56 \textit{K})$ and extrapolate to 1, demonstrating that all Cu spins are observed in the charge-ordered phase. The relaxation changes from Gaussian at high temperatures to purely exponential below $T_{CO}$, similar to previous results on other cuprates. Data have been corrected for the usual Boltzmann factor and line width change. The constant offset in the 42~K curve is the noise floor. (b) spin-lattice relaxation curves for the Cu A line at 45~K, isotope $^{65}$Cu (squares) and $^{63}$Cu (circles). The echo time used in the detection sequence is 2.8~$\mu$s. The relaxation times are indistinguishable within experimental uncertainty, and no significant stretching of the exponential relaxation is observed. The inset shows the magnetization recovery curve with a logarithmic time scale.
}
\end{figure}

Spin-spin and spin-lattice relaxation provide additional insight into the dynamics of the ordered phase, and show dramatic effects close to $T_{CO}$. In Fig. \ref{spins} we show two representative spin-spin relaxation curves for B line Cu spins, demonstrating that wipeout is indeed eliminated if the measurement is performed at short times -- the number of detected spins is the same (within experimental uncertainty) above and below $T_{CO}$. The full temperature dependence of the integrated intensity of the B line, corrected for temperature and $T_2$ \cite{T2correction}, is shown in the inset of Fig. \ref{order}, with no appreciable decrease down to the spin ordering temperature $T_{SO}$. This is evidence that the wipeout observed previously in LBCO and similar compounds is predominantly a spin-spin relaxation effect above $\sim 40$~K; at lower temperatures the spin-lattice relaxation also becomes fast and contributes to wipeout (see below). The shape of the spin-spin relaxation curve changes from Gaussian above $T_{CO}$ to purely exponential deep in the ordered phase, an effect already related to charge order in YBCO [\onlinecite{YBCONMR1}] and 214 cuprates\cite{wipeout4}. The spin-spin relaxation time $T_2$ abruptly decreases below $T_{CO}$, and continues to decrease as the charge order develops. The relevant time for spin echo measurements, $T_2 / 2$, is always below 5~$\mu$s in the ordered phase. The signal is thus virtually unobservable in a standard NMR/NQR experiment\cite{stripe_glass,wipeout_zn} with minimal echo times $\sim 15$~$\mu$s. Wipeout in LBCO is not an exception: similar strong signal decrease below $T_{CO}$is found in Nd-LSCO and LESCO, with the latter showing additional partial wipeout\cite{LESCOmi} above $T_{CO}$. Here we show that the apparent wipeout fraction depends on the NMR/NQR pulse configuration if the experiment is not performed with short echo times; this implies that the wipeout onset temperatures determined in previous work are to some extent dependent on experimental conditions, and could also clarify the incomplete wipeout seen in LESCO. 

\begin{figure}
\centering
\includegraphics[width=85mm]{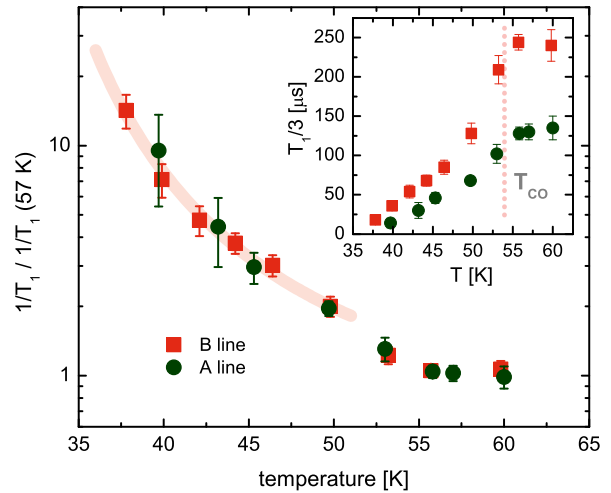}
\caption{\label{T1} (color online) Spin-lattice relaxation rates for A and B lines in dependence on temperature, normalized to the respective high-temperature values. A strong increase of the relaxation rate is observed in the charge-ordered phase. Solid line is a fit to the Vogel-Fulcher-Tamman relation used for glassy relaxation (see text). Inset: raw $T_1$ data for the A and B lines. 
}
\end{figure}

The behaviour of spin-lattice relaxation times $T_1$ is qualitatively similar to $T_2$, with a drop at $T_{CO}$ and strong decrease in the ordered phase (Fig. \ref{T1} inset). However, above $\sim 40$~K the $T_1$ values are an order of magnitude longer than $T_2$, and thus $T_2$ relaxation is the dominant cause of wipeout in that range. Below $T_{SO}$ the spin-lattice relaxation becomes comparable to spin-spin relaxation and contributes to wipeout -- measurements below $\sim 38$~K are impossible in our experimental configuration due to short relaxation times. The spin-lattice relaxation times of A and B lines are the same up to a temperature-independent factor of $\sim 1.8$, which is not a consequence of charge order since it is the same above and below $T_{CO}$. 
At 45~K we measured the relaxation times of the two Cu isotopes at the peaks of the A lines, in order to determine if the relaxation mechanism is quadrupolar or magnetic. The relaxation times are the same within error limits: their ratio is $^{63}T_1 / ^{65}T_1 = 1.04 \pm 0.07$, consistent with magnetic relaxation and to be discussed in more detail below. We note that the spin-lattice relaxation curves were fitted to a pure exponential function $1-M(t)/M(\infty) = e^{-3t/T_1}$ appropriate for spin-3/2 NQR, without introducing an exponential stretching exponent. This increases the reliability of the fits, especially at the lowest temperatures where the relaxation is very fast. As seen in Fig. \ref{spins}, at 45~K the stretching is negligible: a fit with included exponent $1-M(t)/M(\infty) = e^{-\left( 3t/T_1 \right)^{\beta}}$ gives $\beta \approx 0.9$. At the lowest available temperature -- 38~K for the B~line -- a stretched fit gives $\beta \approx 0.7$ and an almost unchanged mean $T_1$, demonstrating that the stretching does not significantly influence the results. 

\section{External magnetic fields}

\begin{figure}[b]
\centering
\includegraphics[width=85mm]{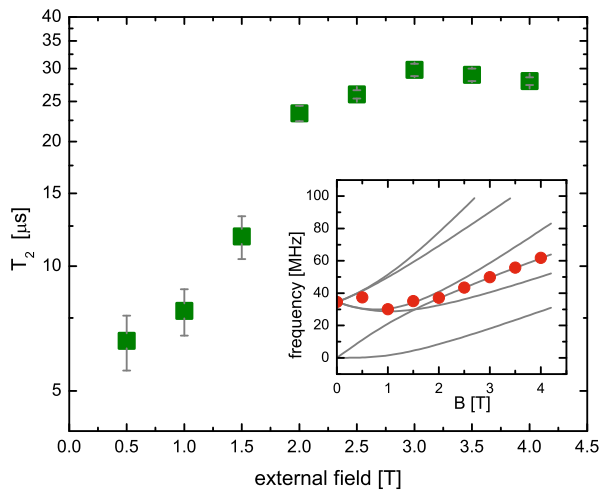}
\caption{\label{T2B} (color online) Cu spin-spin relaxation times at 50~K in dependence of an applied external field in the copper-oxygen plane. The fluctuations responsible for the fast spin-spin relaxation at $B = 0$~T are all but absent above $\sim 2.5$~T. Inset: lines are the six transitions of a $^{63}$Cu nucleus (spin $3/2$) calculated numerically for an external field perpendicular to the EFG symmetry axis (see text), points are actual frequencies used in the experiment to obtain the results in the main figure.
}
\end{figure}

To discuss the effects of applied magnetic fields, we must first consider how a field modifies the NQR/NMR spectrum of copper. Since the relevant magnetic fields are relatively small, we cannot use the usual high-field approximation but have to diagonalize the nuclear Hamiltonian exactly. For a spin-3/2 nucleus such as $^{63}$Cu, the static Hamiltonian reads\cite{Abragam}
\begin{equation}
\label{hamiltonijan}
H=-\gamma \mathbf{B} \mathbf{I} + \frac{e^2QV_{zz}}{4I(2I-1)}\left[
3I_{z}^2-I^2+\eta\left( I_x^2-I_y^2\right)
\right]
\end{equation}
where $\gamma$ is the gyromagnetic ratio of the $^{63}$Cu nucleus,
$\mathbf{I}$ its spin, $Q$ its quadrupole moment, $\mathbf{B}$ the external
magnetic field, $V_{xx}$, $V_{yy}$ and $V_{zz}$ the principal components of
the electric field gradient (EFG) tensor, and $\eta =
\left|\left(V_{xx}-V_{yy}\right) /V_{zz}\right| \approx 0$ in LBCO. Two contributions determine the NMR spectrum: the Zeeman ($B$-dependent) and quadrupolar ($V$-dependent) terms. In our case the Zeeman and quadrupolar terms are generally comparable, and we thus diagonalize the Hamiltonian numerically to obtain the exact transition frequencies in dependence on external field applied in the CuO plane (perpendicular to the axis of the quadrupolar tensor).  The resulting six transitions are plotted in the inset of Fig. \ref{T2B}, together with the actual frequencies employed in the measurements. 

The effect of an in-plane magnetic field on the fluctuations causing Cu spin-spin relaxation is dramatic. The relaxation time $T_2$ increases by a factor $\sim 5$ from pure NQR to 3~T (Fig. \ref{T2B}) at 50~K. But notably the relaxation is still faster than above $T_{CO}$ without external field, and remains pure exponential at 4~T. 

\section{Discussion}

Our investigation of Cu NQR/NMR in the charge-stripe phase of LBCO-1/8 resolves several disputed issues and enables a direct comparison to other cuprates such as YBCO. First, we demonstrate that the wipeout effect can be eliminated if the experiment is performed with echo times short enough. We therefore establish directly that the wipeout is solely due to spin fluctuations, with no direct relation to the charge order which is static on the NQR timescale. Enhanced spin fluctuations generally appear simultaneously with charge stripes in the cuprates\cite{YBCONMR1,LBCOLaNMR,stripe_glass}, and their effect on Cu $T_2$ is about an order of magnitude stronger in LBCO than in YBCO\cite{YBCONMR1}. This is in line with the increased propensity of 214 cuprates to antiferromagnetic (AFM) fluctuations seen at all temperatures\cite{wipeout_UD,LSCOCuNMR1} in NMR. However, the magnetic fluctuations in LBCO are dynamically decoupled from charge order (at least in the temperature range here investigated) -- charge stripe order is static on the NQR scale\cite{CDWstatic}, leading to symmetric line broadening, while magnetic fluctuations cause relaxation rate enhancement. The possibility of a charge fluctuation-related $T_1$ mechanism is eliminated through the measurement of relaxation rates for the two Cu isotopes: for a mechanism coupling to the quadrupolar moment of the nuclei, the ratio of relaxation times should be $^{63}T_1/^{65}T_1 = \left(^{65}Q/^{63}Q \right)^2 \approx 0.85$, while the experimental ratio is $\approx 1$. Yet equal relaxation times are consistent with a standard Bloembergen-Purcell-Pound (BPP) magnetic mechanism with a characteristic spin fluctuation timescale $\tau_c$, leading to\cite{BPP} $1/T_1 \sim \gamma^2 \omega^2 \tau_c$ where $\gamma$ is the gyromagnetic ratio of the nucleus, $\omega$ the resonance frequency, and the simple form is valid when $1/\tau_c \gg \omega$. Since $\omega \sim Q$ in NQR, the BPP formula yields a relaxation time ratio $^{63}T_1/^{65}T_1 = \left(^{65}Q/^{63}Q \right)^2 \left(^{65}\gamma/^{63}\gamma \right)^2 \approx 1.02$. We thus conclude that charge stripes are effectively pinned as soon as they appear, while the spins in the stripes remain slowly fluctuating. The glass-like slowing down of the magnetic fluctuations has been extensively investigated with La NMR, and here we see that their effect is $\sim 1000$ times stronger at the Cu sites. From the $T_1$ data we can estimate the temperature dependence of the relaxation time $\tau_c$ using the BPP formula above: $\tau_c \sim 1/T_1$, Fig. \ref{T1} main graph. The points are consistent with a Vogel-Fulcher-Tammann law\cite{glass} used in the physics of glassy systems,
\begin{equation}
\label{VFT}
\tau_c = \tau_0 e^{\Delta/\left(T-T_{VFT}\right)}
\end{equation}
where $\Delta$ is a characteristic energy scale, and $T_{VFT}$ the effective 'freezing' temperature where $\tau_c$ diverges. A fit of relation (\ref{VFT}) to the data in Fig. \ref{T1} gives $\Delta \approx 80$~K and $T_{VFT} \approx 20$~K (a pure activational dependence would have $T_{VFT} = 0$); while the value of $T_{VFT}$ should be taken with caution due to the limited range of the fit, it is consistent with the temperatures where Cu spin freezing is observed\cite{LSCOmu,LESCOmu} in LSCO and LESCO using $\mu$SR, with the temperature range where $^{139}$La spin-lattice relaxation rates start to decrease\cite{LBCOLaNMR}, and with the temperature where 2D superconductivity is formed through a Berezinski-Kosterlitz-Thouless transition\cite{LBCO2D}. The spin-order temperature $T_{SO} \sim 40$~K detected using neutron scattering\cite{LBCOtransport} is somewhat higher than the freezing scale $T_{VFT}$ most probably due to the different timescales probed by neutrons and NQR. $T_{VFT} > 0$ shows qualitatively that a glassy description of the magnetic dynamics is appropriate. This is relevant to many cuprate systems, where similar slowing down without long-range magnetic order occurs at low temperatures\cite{spinglass1,spinglass2}.

The question of spin-echo decay is also one to be discussed in light of our results. Namely, the crossover from a Gaussian decay in the high-temperature phase to exponential below $T_{CO}$ we see here has been reported for other 214 cuprates\cite{wipeout4} and YBCO\cite{YBCONMR1}, but with somewhat incomplete explanations. It is known that the Gaussian decay in cuprates is generally caused by a strong Cu-Cu transferred coupling\cite{trans1,trans2}, and an unlike-spin coupling has been invoked to explain the crossover to exponential relaxation\cite{wipeout4}. Yet such an explanation requires the existence of a significant fraction of wiped-out, quickly relaxing nuclei which cause the exponential echo decay of the observable nuclei. In our case of Cu NQR this is ruled out, since within experimental uncertainty we detect all nuclei (at least above $\sim 40$~K), and know that the relaxation times are more or less homogeneous. A similar situation occurs in YBCO in the charge-ordered phase\cite{YBCONMR1}. Hence the origin of exponential decay must be different: it is beyond this work to provide a quantitative theory, but we can attempt a qualitative explanation as follows. The spin-spin decay curve will have a Gaussian shape if the transferred coupling between Cu nuclei is essentially static, i.e. when mutual spin flips are rare\cite{trans1,trans2}. This happens in Cu NMR with the magnetic field perpendicular to the CuO$_2$ planes, and in NQR at elevated temperatures. Yet if spin flips cannot be neglected, the spin-spin relaxation curve will change. In the limit where spin-flip processes dominate, it becomes pure exponential. An \emph{in-plane} magnetic field makes spin flips important and causes an exponential relaxation\cite{trans1}; it therefore seems plausible that strong in-plane electron spin fluctuations have a similar effect. The relaxation thus becomes exponential when the spin fluctuations associated with charge stripe order occur below $T_{CO}$. Importantly, all Cu nuclei feel the same influence, in agreement with both YBCO and our LBCO results. We note that a Redfield-like $T_1$ related echo decay\cite{Slichter} cannot be operative in LBCO except perhaps below 40~K, since at higher temperatures the $T_1$ values are still too long to cause a $T_2$ as short as $\sim 5$~$\mu$s.

The effects of external magnetic field are also consistent with recent results on $^{139}$La, but again much more pronounced\cite{LBCOLaNMR}. The high NQR frequency of Cu enables us to study low fields, revealing the characteristic fluctuation suppression scale of $\sim 2$~T for in-plane fields. In the relevant temperature range the Cu spins in LBCO belong to the 2D XY universality class\cite{LBCOtransport,LBCOLaNMR}, implying that an external field suppresses AFM fluctuations by creating a preferred direction in the planes\cite{LBCOLaNMR}. Our results are not directly comparable to $^{139}$La NMR (where only data above 5~T and at 24~K are available), but the same tendency is observed there. Importantly, the in-plane field effectively decreases the spin-spin relaxation rate and thus modifies wipeout, in agreement with Cu NQR wipeout measurements on LESCO-1/8 crystals\cite{LESCOmi}. All these observations are a consequence of the same suppression effect, and our Cu measurements here show that it is still present at fields as low as $\sim 2$~T in the charge-ordered phase above the spin-ordering temperature $T_{SO} \approx 40$~K.

We see no significant differences between A and B sites in either relaxations or static NQR lineshapes, demonstrating (i) that the magnetic fluctuations are insensitive to barium impurities and (ii) that charge stripes are not pinned to the Ba local fields. The latter conclusion agrees with scattering studies, which show that stripes are pinned to orthorhombic twin boundaries\cite{COxray2} in LSCO. Yet in contrast to scattering methods, the Cu NQR lineshape provides a microscopic measure of the amplitude of the charge modulation. The relative linewidth change of the Cu B line in LBCO at 40~K is $\Delta\nu_Q / \nu_Q \sim 0.012$, and can be extrapolated to $\sim 0.02$ at 0~K (assuming that the linewidth follows the X-ray charge order parameter in the entire temperature range). This is about 2 times larger than the charge-stripe induced Cu NMR line splitting\cite{YBCONMR1} in YBCO-0.108 -- the values are probably even closer at doping 1/8 where the splitting in YBCO is expected to be the largest\cite{COxray1,COxray5,YBCONMR1}, but a direct comparison to YBCO-1/8 is not available due to the complicated Cu NMR spectra. Furthermore, the Cu linewidth change in LBCO is similar to the $^{17}$O quadrupolar line broadening/shift in YBCO-1/8 [\onlinecite{YBCONMR3}] and LESCO [\onlinecite{Owipeout}]. This indicates that the amplitude of charge modulation in the cuprates is universal, i.e. does not significantly depend on the material family. A similar conclusion was reached by resonant soft x-ray diffractionp{universalx}. In contrast, the accompanying magnetic fluctuations are strongly family-dependent, suggesting that charge order is an intrinsic feature of the CuO$_2$ planes around 1/8 doping, with the strength of associated spin fluctuations determined by the underlying AFM tendencies of the family.

\section{Summary}

To conclude, we present the first Cu NQR/NMR investigation of charge-stripe ordered LBCO-1/8 which eliminates the wipeout effect, thus enabling insight into the microscopic physics of charge stripes and associated magnetic fluctuations. We find that the charge stripes are static on the NMR scale $\sim 0.1$~$\mu$s as soon as they appear, and that the charge modulation amplitude can be reliably determined from Cu NQR linewidth. A comparison to results on other cuprates indicates that the relative amplitude is universal in the cuprates. Strong spin fluctuations accompany the charge order, and evidence for glassy spin dynamics/spin freezing is obtained from Cu nuclear spin relaxation measurements, corroborating previous $^{139}$La NMR studies. The magnetic fluctuations also cause the apparent Cu signal wipeout by making the relaxation times significantly shorter than the typical NMR measurement times. Our fast spin echo experiment provides a way to detect the quickly-relaxing nuclei and avoid wipeout, and could thus be relevant for measuring previously unseen NMR/NQR signals in a wide class of materials where strong spin fluctuations appear.

\acknowledgements{
Discussions with J. M. Tranquada, A. Dulčić and M. Grbić and the assistance of B. Mihaljević in NQR experiments are gratefully acknowledged. D. P. and M. P. acknowledge funding by the Croatian Science Foundation (HRZZ) under grant no. IP-11-2013-2729. The work at Brookhaven National Laboratory was supported by the Office of Basic Energy Sciences, U.S. Department of Energy, under Contract No. DE-SC00112704.
}


\begin{thebibliography}{1}

\bibitem{LBCOneutron} M. Hücker et al., \emph{Phys. Rev. B} \textbf{83,} 104506 (2011)

\bibitem{LBCO2D} Q. Li, M. Hücker, G. D. Gu, A. M. Tsvelik, J. M. Tranquada, \emph{Phys. Rev. Lett.} \textbf{99}, 067001 (2007)

\bibitem{COxray1} J. Chang et al., \emph{Nature Phys.} \textbf{8}, 871 (2012)

\bibitem{COxray2} J. Fink et al., \emph{Phys. Rev. B} \textbf{83,} 092503 (2011)

\bibitem{COxray3} N. B. Christensen et al., \emph{arXiv:1404.3192} (2014)

\bibitem{COxray4} W. Tabis et al., \textit{Nature Comm.} \textbf{5}, 5875 (2014)

\bibitem{COxray5} M. H\"ucker et al., \emph{Phys. Rev. B} \textbf{90}, 054514 (2014)

\bibitem{YBCONMR1} T. Wu et al., \emph{Nature} \textbf{477}, 191 (2011)

\bibitem{YBCONMR2} T. Wu et al., \emph{Nature Comm.} \textbf{4}, 2113 (2013)

\bibitem{YBCONMR3} T. Wu et al., \emph{Nature Comm.} \textbf{6}, 7438 (2015)

\bibitem{LSCOLaNMR1} S.-H. Baek, A. Erb, B. Büchner, H.-J. Grafe, \textit{Phys. Rev. B} \textbf{85}, 184508 (2012)

\bibitem{LSCOLaNMR2} V. Mitrović et al., \emph{Phys. Rev. B} \textbf{78}, 014504 (2008)

\bibitem{LBCOLaNMR} S.-H. Baek et al., \emph{Phys. Rev. B} \textbf{92,} 155144 (2015)

\bibitem{stripe_glass} A. W. Hunt, P. M. Singer, A. F. Cederstr\"om and T. Imai, \emph{Phys. Rev. B} \textbf{64,} 134525 (2001)

\bibitem{LESCOLaNMR} S.-H. Baek et al., \textit{Phys. Rev. B} \textbf{87}, 174505 (2013)

\bibitem{wipeout1} A. W. Hunt, P. M. Singer, K. R. Thurber and T. Imai, \textit{Phys. Rev. Lett.} \textbf{82}, 4300 (1999)

\bibitem{wipeout2} P. M. Singer, A. W. Hunt, A. F. Cederstr\"om and T. Imai, \emph{Phys. Rev. B} \textbf{60,} 15345 (1999)

\bibitem{wipeout3} H.-J. Grafe et al., \emph{Eur. Phys. J. ST} \textbf{188,} 89-101 (2010)

\bibitem{Owipeout} H.-J. Grafe, N. J. Curro, M. H\"ucker and B. B\"uchner, \emph{Phys. Rev. Lett.} \textbf{96,} 017002 (2006)

\bibitem{wipeout4} N. J. Curro et al., \emph{Phys. Rev. Lett} \textbf{85,} 642-645 (2000)

\bibitem{LESCOmi} D. Pelc et al., \emph{Nature Comm.} \textbf{7}, 12775 (2016)

\bibitem{wipeoutspin} M.-H. Julien et al., \emph{Phys. Rev. B} \textbf{63,} 144508 (2001)

\bibitem{wipeout_zn} H.-J. Grafe et al., \textit{Phys. Rev. B} \textbf{77}, 014522 (2008)

\bibitem{wipeout_UD} S. Ohsugi et al., \textit{J. Phys. Soc. Jpn.} \textbf{63}, 700 (1994)

\bibitem{damp1} D. I. Hoult, \textit{Rev. Sci. Instrum.} \textbf{50}, 193 (1979)

\bibitem{damp2} A. S. Peshkovsky, J. Forguez, L. Cerioni and D. J. Pusiola, \textit{J. Magn. Reson.} \textbf{177}, 67 (2005)

\bibitem{fukushima} E. Fukushima and S. Roeder, \textit{Experimental Pulse NMR.} (Westview Press, Boulder, 1981)

\bibitem{LBCOtransport} J. M. Tranquada et al. \emph{Phys. Rev. B} \textbf{78,} 174529 (2008)

\bibitem{T2correction} The spin-spin relaxation correction was done using a Gaussian relaxation curve above 50~K, a pure exponential below 47~K, and an interpolation between the two at 49~K.

\bibitem{214disorder} P. M. Singer, A. W. Hunt and T. Imai, \textit{Phys. Rev. Lett.} \textbf{88}, 047602 (2002)

\bibitem{Abragam} A. Abragam, \emph{Principles of Nuclear Magnetism.} (Oxford Univ. Press, Oxford, 1983)

\bibitem{LSCOCuNMR1} T. Imai, C. P. Slichter, K. Yoshimura and K. Kosuge, \textit{Phys. Rev. Lett.} \textbf{70}, 1002 (1993)

\bibitem{CDWstatic} X. M. Chen et al., \textit{arxiv:1606.04168} (2016)

\bibitem{BPP} N. Bloembergen, E. M. Purcell and R. V. Pound, \emph{Phys. Rev.} \textbf{73}, 679 (1948)

\bibitem{glass} E. Donth, \textit{The Glass Transition.} (Springer, Berlin, 2001)

\bibitem{LSCOmu} A. T. Savici et al., \emph{Phys. Rev. B}  \textbf{66}, 014524 (2002)

\bibitem{LESCOmu} H.-H. Klauss et al., \emph{Phys. Rev. Lett.} \textbf{85,} 4590 (2000)

\bibitem{trans1} C. H. Pennington et al., \emph{Phys. Rev. B} \textbf{39}, 274 (1989)

\bibitem{trans2} C. H. Pennington and C. P. Slichter, \emph{Phys. Rev. Lett.} \textbf{66}, 381 (1991)

\bibitem{Slichter} C. P. Slichter, \textit{Principles of Magnetic Resonance.} (Springer, Berlin, 1996)

\bibitem{spinglass1} S. Sanna et al., \textit{Phys. Rev. Lett.} \textbf{93}, 207001 (2004) 

\bibitem{spinglass2} H.-H. Klauss, \textit{J. Phys.: Cond. Mat.} \textbf{16}, S4457 (2003)

\bibitem{universalx} V. Thampy et al., \textit{Phys. Rev. B} \textbf{88}, 024505 (2013) 

\end{thebibliography}
\end{document}